\documentclass[11pt]{article}
\usepackage{amssymb,amsmath,cite,geometry,graphicx,url}
\catcode`\@=11

\@addtoreset{equation}{section}
\def\theequation{\arabic{section}.\arabic{equation}}
     
\catcode`\@=11
\def\thesection{\arabic{section}.}

\def\appendix{\setcounter{section}{0}
        \def\thesection{Appendix.}
        \def\theequation{\Alph{section}.\arabic{equation}}}
\def\section{\@startsection{section}{1}{\z@}{3.5ex plus 1ex minus
   .2ex}{2.3ex plus .2ex}{\large\bf}}

\long\def\@makefntext#1{\parindent 0cm\noindent
\hbox to 1em{\hss$^{\@thefnmark}$}#1}

\newcommand{\captionfonts}{\small}
\makeatletter  
\long\def\@makecaption#1#2{%
  \vskip\abovecaptionskip
  \sbox\@tempboxa{{\captionfonts #1: #2}}%
  \ifdim \wd\@tempboxa >\hsize
    {\captionfonts #1: #2\par}
  \else
    \hbox to\hsize{\hfil\box\@tempboxa\hfil}%
  \fi
  \vskip\belowcaptionskip}
\makeatother   

\begin{document}
\begin{titlepage}
\vspace{.5in}
\begin{flushright}
 
June 2015\\  
\end{flushright}
\vspace{.5in}
\begin{center}
{\Large\bf
 Dimensional reduction in causal set gravity}\\  
\vspace{.4in}
{S.~C{\sc arlip}\footnote{\it email: carlip@physics.ucdavis.edu}\\
       {\small\it Department of Physics}\\
       {\small\it University of California}\\
       {\small\it Davis, CA 95616}\\{\small\it USA}}
\end{center}

\vspace{.5in}
\begin{center}
{\large\bf Abstract}
\end{center}
\begin{center}
\begin{minipage}{4.5in}
{\small
Results from a number of different approaches to quantum gravity suggest 
that the effective dimension of spacetime may drop to $d=2$ at small scales.
I show that two different dimensional estimators in causal set theory display 
the same behavior, and argue that a third, the spectral dimension, may 
exhibit a related phenomenon of ``asymptotic silence.''
}
\end{minipage}
\end{center}
\end{titlepage}
\addtocounter{footnote}{-1}

\section{Introduction} 

In classical physics, the dimension of spacetime is a fixed parameter,
specified from the outset.  In quantum gravity, this may no longer be the 
case: dimension may be a quantum observable, taking different values under 
different circumstances.  In particular, there are intriguing hints from a number 
of different approaches to quantum gravity that the effective dimension of 
spacetime drops to two at very small distances \cite{Carlip1,Carlip2}.  This 
``spontaneous dimensional reduction'' was first noted in high temperature 
string theory \cite{Atick}, and then a few years later in the discrete path 
integral approach of causal dynamical triangulations \cite{Ambjorn}.  
Since then, the same behavior has been seen in the asymptotic safety 
program \cite{Reuter,Percacci}, the short distance approximation to the 
Wheeler-DeWitt equation \cite{Carlip1}, aspects of loop quantum gravity 
\cite{Modesto}, some formulations of noncommutative geometry 
\cite{Benedetti,Nozari,Arzano} or minimum length \cite{Modestob},
and perhaps Ho{\v r}ava-Lifshitz gravity \cite{Horava}.  The generality 
of these results suggests that dimensional reduction may be a fundamental 
feature of quantum gravity.

One important approach to quantum gravity, however, seems to be an 
exception.  In \cite{Eichhorn}, Eichhorn and Mizera show that the spectral 
dimension of a causal set \emph{increases} at short distances.  In this paper, 
I will show that two other dimensional estimators for causal sets---the
Myrheim-Meyer dimension of a small causal set and the dimension
determined by the causal set Laplacian---display the more standard drop
to $d=2$ and short distances.  I will argue that the Eichhorn--Mizera result
may have a different interpretation, as an indication of short distance
``asymptotic silence'' \cite{Heinzle}, a behavior that has also been
associated with dimensional reduction \cite{Carlip1,Carlip2,Carlip3}.

This paper should be read as a report on work in progress.  As we
shall see, a number of relevant concepts (e.g., Hadamard Greens
functions on causal sets) and calculations (e.g., Myrheim-Meyer
dimension for ``small'' causal sets with more than six elements) do
not yet exist.  But the preliminary results are promising, and this seems 
to be a program worthy of further study.

\section{Causal sets}

A causal set \cite{Bombelli} is a discrete spacetime in which events 
have prescribed causal relations.  Such a set is characterized by a partial 
order $\prec$ (where $x\prec y$ means ``$x$ is to the past
of $y$'') satisfying\\[-4ex]
\begin{enumerate}\addtolength{\itemsep}{-1.5ex}
\item transitivity: $x\prec y$ and $y\prec z \Rightarrow x\prec z$;
\item acyclicity: $x \prec y$ and $y \prec x \Rightarrow x=y$;
\item local finiteness: for any $x$ and $y$, the number of elements
   $z$ such that $x\prec z\prec y$ is finite.
\end{enumerate}
Mathematically, these conditions define a locally finite partially ordered
set, or poset.  Physically, the causal relations should be thought of as
determining ``most'' of the metric.  Indeed, in the continuum, the causal 
structure of a globally hyperbolic manifold determines the metric 
up to a conformal factor \cite{Malamet}; in causal set theory, the
missing conformal factor is simply the number of points in a region.

Causal sets with clear physical meaning can be constructed
by randomly ``sprinkling'' points on a fixed spacetime.
Given a globally hyperbolic manifold $M$ with metric $g$, select 
a set of points by a Poisson process so that the probability of 
finding $m$ points in any region of volume $V$ is 
\begin{align}
P_V(m) = \frac{(\rho V)^m}{m!}e^{-\rho V}
\label{a1}
\end{align}
for some discreteness scale $\rho^{-1}$.  Assign to these points the
causal relations determined by the metric $g$, and then ``remove''
the manifold $M$, leaving only the set of points and relations.   At 
scales larger than $\rho^{-1}$, the resulting causal set is believed to 
approximate $M$ well.  In particular, if $M$ is Minkowski space, the 
causal set preserves statistical Lorentz invariance \cite{LIV}, a highly 
nontrivial characteristic for any discretization.

\section{Myrheim-Meyer dimension of a small causal set}

As in other discrete approaches to quantum gravity, it is not obvious
what one means by the ``dimension'' of a causal set.  For a space with
an analog of a Riemannian metric, a popular choice is the spectral
dimension, but it is not obvious that this is appropriate to a Lorentzian 
spacetime; I will return to this issue later.  For a causal set, the most
common choice for a dimensional estimator is the Myrheim-Meyer 
dimension \cite{Myrheim,Meyer}, which is based on a count of the 
number of causally related points.

More precisely, let us start with a causal set derived from a Poisson 
sprinkling of points in $d$-dimensional Minkowski space.  Choose 
an Alexandrov interval, or ``causal diamond,'' $\mathcal{A}$, that 
is, the intersection of the future of some point $p$ and the past 
of another point $q$.  Let $\langle C_1\rangle$ be the average
number of points in $\mathcal{A}$, and $\langle C_2\rangle$
be the average number of causal relations, that is, pairs $x,y$ such 
that $x\prec y$.  $\langle C_1\rangle$  and $\langle C_2\rangle$
depend on the volume of $\mathcal{A}$ and the discreteness scale
$\rho^{-1}$, but a suitable ratio depends only on the dimension:
\begin{align}
\frac{\langle C_2\rangle\,}{\langle C_1\rangle^2} =
   \frac{\Gamma(d+1)\Gamma(\frac{d}{2})}{4\Gamma(\frac{3d}{2})}
\label{b1}
\end{align}
For an arbitrary causal set, the Myrheim-Meyer dimension $d_M$ 
is then defined as the value $d$ for which (\ref{b1}) holds.  One can
also consider a sprinkling of points in a curved spacetime; if the
curvature is small, a generalization of (\ref{b1}) involving chains of
three and four related points can eliminate distortions due to
curvature \cite{Roy}.

We are interested here in the dimension of ``small'' causal sets.  
There are several different things this might mean:\\[-4ex]
\begin{enumerate}\addtolength{\itemsep}{-1.5ex}
\item One might simply take a random causal set with a small
number $C_1$ of elements.  As a practical matter, $C_1$ must be 
\emph{very} small: the number of distinct causal sets with $C_1$ 
elements goes as $2^{C_1^2/4}$, and the causal sets have
only been fully enumerated up to $C_1=16$ \cite{McKay}.
\item For larger $C_1$, random causal sets are dominated by 
Kleitman-Rothschild, or KR, orders \cite{KR,Henson}.  These are
three-layered posets with approximately $C_1/4$ elements in
the first and third layers and $C_1/2$ elements in the second;
an element in the first or third layer is causally related 
to about half of the  elements in the second layer, and almost 
every element in the first layer is related to almost every 
element in the third.  Numerical studies indicate that these  
sets become important at $C_1\sim 50$ \cite{Henson}.  While 
KR orders must be dynamically suppressed at large scales if causal 
set theory is to reproduce anything like our universe, 
it is plausible that they remain important at reasonably small 
scales.
\item The preceding criteria do not include dynamics,
in part because the dynamical behavior of causal set theory is 
not well understood.  One might, however, consider  
random sprinklings of points in known spacetimes---Minkowski 
space, for instance---and look at their small scale behavior.
\end{enumerate} 

The first two of these approaches show clear signs of 
dimensional reduction.  For example, suppose we start with a 
large causal set and chose a subset containing four elements.
There are a total of 16 possible causal structures among those elements,
having between zero and six causal relations.  If these structures
occur with equal probability, the average $\langle C_2\rangle$
is $\frac{13}{4}$, and the Myrheim-Meyer dimension (\ref{b1})
is 2.27.   For random causal sets with four, five, or six 
elements, as enumerated in the Chapel Hill poset atlas \cite{CH},
the Myrheim-Meyer dimensions range from 2.15 to 2.27.
Similarly, for a random KR order, the dimension is 2.38.  

For the third approach, more numerical work is needed.  But 
Reid has looked at random sprinklings in Minkowski space 
\cite{Reid}, and the results show a decrease in the Myrheim-Meyer 
dimension to a bit less than 2 for small subintervals, as expected
in short distance dimensional reduction.

\section{Laplacians and Greens functions}

Consider a massless field in a $d$-dimensional spacetime.  At
short distances, the Hadamard Greens function takes the form 
\begin{align}
G^{(1)}(x,x') \sim \left\{ \begin{array}{lc} \sigma(x,x')^{-(d-2)/2} \quad& d>2\\
                              \ln\sigma(x,x') & d=2 \end{array}\right.
\label{c1}
\end{align}
where Synge's world function $\sigma(x,x')$ is half the squared
geodesic distance between $x$ and $x'$.  The dimension is thus 
determined, in a manifestly physical way, by the rate at which 
the two-point function blows up at coincident points.

As usual, it is not immediately obvious how to extend this expression 
to a discrete spacetime.  Recently, however, considerable progress has 
been made in defining Laplacians and retarded Greens functions 
on causal sets obtained by random sprinklings of points in Minkowski 
space in two \cite{Sorkina}, four \cite{Benincasa}, and arbitrary 
\cite{Dowker,Aslanbeigi} dimensions.  While the retarded Greens
functions are not the same as the Hadamard functions (\ref{c1}),
they still provide use useful information.

Aslanbeigi et al.\ have examined the behavior of these quantities 
averaged over causal sets obtained by sprinklings on $d$-dimensional
Minkowski space \cite{Aslanbeigi}.  The averaged Laplacians have 
plane wave eigenfunctions $e^{ip\cdot x}$, as expected from Poincar{\'e}
invariance, with calculable eigenvalues $g(p)$.  Hence
\begin{align}
G_R(x,x') = \int_{\mathcal{C}}\!d^dp\, g(p)^{-1} e^{ip\cdot(x-x')}
\label{c1a}
\end{align}
In the IR limit relevant for long distance behavior, 
$g(p)^{-1}\sim 1/p^2$,  confirming that the 
causal set Laplacians approximate the standard continuum
operators.  In the UV, though, one finds that
\begin{align}
g(p)^{-1} \sim \alpha + \beta (p\cdot p)^{-d/2}
\label{c2}
\end{align}
For the contour $\mathcal{C}$ appropriate for a retarded Greens function, 
the integral (\ref{c1a}) near the coincidence limit $\sigma\rightarrow0$
gives a delta function plus a finite correction, the normal behavior 
for a retarded Greens function \cite{Aslanbeigi}.  But if, as in the 
continuum case, the Hadamard function can be obtained by choosing a 
different contour in (\ref{c1a}), then  (\ref{c2}) will lead to a Hadamard 
function $G^{(1)}\sim\ln\sigma$ at short distances, the standard form 
for a two-dimensional massless field theory,\footnote{A similar phenomenon 
occurs in Ho{\v r}ava-Lifshitz gravity \cite{Horava}, but in contrast to that 
model, the causal set result does not require a violation of Lorentz invariance.}  
although one may worry whether this reduction occurs below the discreteness 
scale.

To be confident of this claim, one would have to construct the full 
analog of the Hadamard Greens function in causal set theory and 
examine its UV limit.  Recent work on field theory on causal sets
\cite{Johnston,Belenchiab} suggests an approach to this problem, 
and work is in progress.   But as in the preceding section, we already 
see strong hints of dimensional reduction.

\section{Spectral dimension and asymptotic silence}

Consider a random walk on a $d$-dimensional manifold with a
Riemannian metric.  Diffusion from an initial position $x$ to a
final position $x'$ in a time $s$ is described by a heat kernel
$K(x,x';s)$, which behaves for small $s$ as \cite{Ambjorn}
\begin{align}
K(x,x';s) \sim (4\pi s)^{-d/2} e^{-\sigma(x,x')/2s}
    \left( 1 + \mathcal{O}(s)\right)
\label{d1}
\end{align}
In particular, the return probability $K(x,x;s)$ is determined by
the dimension.  By generalizing (\ref{d1}) to an arbitrary space,
discrete or continuous, on which a random walk can be defined,
one obtains an effective dimension, the spectral dimension.

For several approaches to quantum gravity, including causal
dynamical triangulations \cite{Ambjorn} and asymptotic safety 
\cite{Reuter}, the spectral dimension exhibits short distance
dimensional reduction to $d=2$.  For causal set theory, though,
it does not.  On the contrary, the spectral dimension increases
at short distances \cite{Eichhorn}.  What should one make of
this?

Eichhorn and Mizera argue in \cite{Eichhorn} that the peculiar
behavior of causal set theory comes from the Lorentzian 
signature of the metric, which in many cases leads to a 
``radical nonlocality''---a typical point can have infinitely
many nearest neighbors, points connected by a single causal
link.  Now, as stressed in \cite{Carlip1,Carlip2}, the importance 
of spectral dimension comes in part from the fact that Greens
functions can be obtained as Laplace transforms of the heat
kernel: (\ref{c1}) is a Laplace transform of (\ref{d1}).  But for
causal sets, the Greens functions of \cite{Sorkina,Benincasa,%
Dowker,Aslanbeigi} contain nonlocal corrections, and the  
direct connection to the heat kernel for a random walk 
may be broken.

The results of \cite{Eichhorn} could, however, have a different
implication.  In a Lorentzian setting, a high spectral dimension%
---especially a high value of the ``causal spectral dimension''
of \cite{Eichhorn}---implies a suppression of the probability
that two random walkers will meet within a given diffusion time.
The observed rapid rise in spectral dimension at very short
distances thus suggests that ``nearby'' points are increasingly
causally disconnected.  A very similar behavior 
occurs in cosmology near a spacelike singularity, where it is  
known as ``asymptotic silence'' \cite{Heinzle}.  As I first pointed
out in \cite{Carlip1}, this phenomenon, which leads to locally
Kasner-like behavior of the metric, might explain dimensional
reduction: at certain scales, $d$-dimensional Kasner space  
has an effective dimension of two \cite{Hu}.

It should be possible to test this conjecture more directly.  In the 
continuum, asymptotic silence is an ``anti-Newtonian'' 
limit, in which the speed of light goes to zero and nearby spacelike 
separated points become (nearly) causally disconnected.  In the 
causal set context, defining ``nearby'' is nontrivial, but not 
impossible \cite{Rideout}, and one can measure the minimum number 
of links $N$  required for two nearby points to share a common point 
in the future.  The short distance asymptotic silence conjecture is 
that while $N$ should behave classically for pairs of points with 
large spatial separations, it should become much larger than its
classical value as the spatial distance shrinks.  If this is 
the case, the arguments of \cite{Carlip1} would again predict 
spontaneous dimensional reduction.

\section{Conclusion}

While the evidence for short distance dimensional reduction in
quantum gravity is far from conclusive, there are enough hints
from enough different approaches to make the phenomenon at 
least plausible.  But details remain elusive.  We do not even 
know whether dimensional reduction is mainly kinematical or 
whether it depends sensitively on the dynamics: in the asymptotic 
safety scenario, for instance, the mere existence of a non-Gaussian 
UV fixed point is enough to indicate two-dimensional behavior 
\cite{Percacci}, while in some approaches based on noncommutative 
geometry the nature of dimensional reduction depends
sensitively on a choice of deformed Laplacian \cite{Arzano}.

Causal set theory offers a promising avenue to explore these
issues.  Much of what we know about causal sets is nondynamical,
and there are several approaches to the dynamics that may not
be equivalent \cite{Wallden}.  While this paper is a start, there
is clearly much more to be done:\\[-4ex]
\begin{itemize}\renewcommand{\labelitemi}{\labelitemii}%
   \addtolength{\itemsep}{-1.5ex}
\item A systematic study of the Myrheim-Meyer dimension of
small subsets of random sprinklings on various known manifolds
could reveal more about the influence of large scale spacetime 
geometry, and thus dynamics, on small scale dimension.  One
might also look at the curvature-corrected dimensional estimator 
introduced in \cite{Roy}.
\item A construction of the causal set Hadamard function, 
perhaps following \cite{Johnston,Belenchiab}, and a study of its 
asymptotics in the manner of \cite{Aslanbeigi}, would tell more reliably
whether Greens functions exhibit dimensional reduction.  One
might also compute the heat kernels, and through that the 
spectral dimensions, of the Laplacians in \cite{Aslanbeigi}.%
\footnote{Just after this preprint first appeared, a
preprint by Belechia et al.\ \cite{Belenchia} answered this
last question.  The heat kernels of the nonlocal Laplacians 
\cite{Aslanbeigi} do, in fact, lead to a spectral dimension that
falls to $d=2$ at short distances.}
\item More direct tests of short distance asymptotic 
silence would certainly be illuminating.
\end{itemize}
It may be that different dimensional estimators  
give different answers, and the full picture might require a better
understanding of the quantum dynamics of causal sets.  But the
preliminary indications of short distance dimensional reduction
in causal set theory seem promising.
 
\vspace{1.5ex}
\begin{flushleft}
\large\bf Acknowledgments
\end{flushleft}

This work was supported in part by Department of Energy grant
DE-FG02-91ER40674.

\end{document}